\documentclass[11pt]{article}

\usepackage{latexsym}
\usepackage{graphicx}
\usepackage{amssymb,amsmath}

\def\mt{{\ifmmode M^{eff}_T\else $M^{eff}_T$\fi}}

\def\e{\epsilon}

\def\e3{$\epsilon_3$}

\def\ch2{$\chi^2$}

\def\co#1{{\ifmmode{\cal O}_{#1}\else${\cal O}_{#1}$\fi}}

\newdimen\unit
\def\point#1 #2 #3{\vbox to0pt{\kern-#2\unit
  \hbox{\kern#1\unit#3}\vss}
 \nointerlineskip}

\newcommand{\be}{\begin{equation}}
\newcommand{\ee}{\end{equation}}
\newcommand{\bea}{\begin{eqnarray}}
\newcommand{\eea}{\end{eqnarray}}

\newcommand{\mev}{\mbox{ MeV}}
\newcommand{\gev}{\mbox{ GeV}}
\newcommand{\tev}{\mbox{ TeV}}

\newcommand{\alphaemmz}{\alpha_{\text{em}}(M_Z)^{\overline{MS}}}
\newcommand{\alphas}{\alpha_s(M_Z)^{\overline{MS}}}

\newcount\hour
\newcount\minute
\newtoks\amorpm
\hour=\time\divide\hour by60 \minute=\time{\multiply\hour by60
\global\advance\minute by- \hour}
\edef\standardtime{{\ifnum\hour<12 \global\amorpm={am}%
    \else\global\amorpm={pm}\advance\hour by-12 \fi
    \ifnum\hour=0 \hour=12 \fi
    \number\hour:\ifnum\minute<100\fi\number\minute\the\amorpm}}
\edef\militarytime{\number\hour:\ifnum\minute<100\fi\number\minute}
\def\bold#1{\setbox0=\hbox{$#1$}%
     \kern-.025em\copy0\kern-\wd0
     \kern.05em\copy0\kern-\wd0
     \kern-.025em\raise.0433em\box0 }

\newcommand{\newc}{\newcommand}
\newc\eg{{\rm {e.g.}}}  \newc\etal{{\rm {et al.}}} \newc\ie{{\rm i.e.}}
\newc\etc{{\rm {etc}}}
\newcommand\lsim{\mathrel{\rlap{\lower4pt\hbox{\hskip1pt$\sim$}}
    \raise1pt\hbox{$<$}}}
\newcommand\gsim{\mathrel{\rlap{\lower4pt\hbox{\hskip1pt$\sim$}}
    \raise1pt\hbox{$>$}}}
\newc{\mhalf}{m_{1/2}}      \newc{\mzero}{m_0}
\newc{\tanb}{\tan\beta}
\newc{\azero}{A_0}
\newc{\at}{A_t} \newc{\ab}{A_b} \newc{\atau}{A_\tau}
\newc{\bmu}{B\mu}           \newc{\sgn}{{\rm sgn}}
\newc{\mone}{M_1}           \newc{\mtwo}{M_2}

\newc{\charone}{\chi_1^\pm} \newc{\mcharone}{m_{\chi_1^\pm}}

\newc{\hl}{h}               \newc{\mhl}{m_{\hl}}   \newc{\gammahl}{\Gamma_{\hl}}
\newc{\hh}{H}               \newc{\mhh}{m_{\hh}}   \newc{\gammahh}{\Gamma_{\hh}}
\newc{\ha}{A}               \newc{\mha}{m_{\ha}}   \newc{\gammaha}{\Gamma_{\ha}}
\newc{\hpm}{H^{\pm}}        \newc{\mhpm}{m_{\hpm}} \newc{\gammahpm}{\Gamma_{\hpm}}
\newc{\hp}{H^{+}} \newc{\mhp}{m_{\hp}} \newc{\hm}{H^{-}}
\newc{\mhm}{m_{\hm}}
\newc{\xt}{X_{t}}           \newc{\xb}{X_{b}}

\newc{\qzero}{Q_0}          \newc{\qstop}{Q_{\widetilde t}}
\newc{\amu}{a_{\mu}}        \newc{\amususy}{a_{\mu}^{\text{SUSY}}}
\newc{\amuexpt}{a_{\mu}^{\text{expt}}}        \newc{\amusm}{a_{\mu}^{\text{SM}}}
\newc{\deltaamususy}{\delta a_{\mu}^{\text{SUSY}}}
\newc\gmtwo{(g-2)_{\mu}} \newc\deltaamu{\Delta a_{\mu}}
\newc{\msbar}{\overline{MS}} \newc{\drbar}{\overline{DR}}
\newc{\yt}{h_t} \newc{\yb}{h_b} \newc{\ytau}{h_{\tau}}

\newc{\mtop}{m_t}               \newc{\mtpole}{M_t}
\newc{\mtaupole}{m_{\tau}^{\text{pole}}}
\newc{\mtmtsmmsbar}{m_t(m_t)^{\msbar}_{{\text{SM}}}}
\newc{\mtmtsmdrbar}{m_t(m_t)^{\drbar}_{{\text{SM}}}}
\newc{\mtmtmssmdrbar}{m_t(m_t)^{\drbar}_{{\text{SUSY}}}}

\newc{\mbmbmsbar}{m_b(m_b)^{\msbar} }

\newc{\mbmbsmmsbar}{m_b(m_b)^{\msbar}_{{\text{SM}}}}
\newc{\mbmzsmmsbar}{m_b(\mz)^{\msbar}_{{\text{SM}}}}
\newc{\mbmzsmdrbar}{m_b(\mz)^{\drbar}_{{\text{SM}}}}
\newc{\mbmzmssmdrbar}{m_b(\mz)^{\drbar}_{{\text{SUSY}}}}

\newc{\mtaumzsmmsbar}{m_{\tau}(\mz)^{\msbar}_{{\text{SM}}}}
\newc{\mtaumzsmdrbar}{m_{\tau}(\mz)^{\drbar}_{{\text{SM}}}}
\newc{\mtaumzmssmdrbar}{m_{\tau}(\mz)^{\drbar}_{{\text{SUSY}}}}

\newc{\mgut}{M_{\rm GUT}}
\newc{\mplanck}{M_{\rm P}}      \newc{\mpl}{M_{\text{Pl}}}
\newc{\msusy}{M_{\rm SUSY}}      \newc{\ms}{M_{\text{S}}}
\newc{\jxf}{J({\xf})}
\newc{\jxfexact}{J_{\rm exact}({\xf})}  \newc{\jxfexp}{J_{\rm exp}({\xf})}
\newc{\VEV}[1]{\langle #1 \rangle}
\newc{\xf}{x_f}
\newc\vrel{v_{\rm rel}}
\newcommand\mchi{m_{\chi}}              
\newc\sell{{\widetilde e}_L}      \newc\msell{m_{\sell}}
\newc\selr{{\widetilde e}_R}      \newc\mselr{m_{\selr}}
\newc\snue{{\widetilde \nu}_e}      \newc\msnue{m_{\snue}}
\newc\snutau{{\widetilde \nu}_\tau}      \newc\msnutau{m_{\snutau}}
\newc\supl{{\widetilde u}_L}      \newc\msupl{m_{\supl}}
\newc\supr{{\widetilde u}_R}      \newc\msupr{m_{\supr}}
\newc\sdl{{\widetilde d}_L}      \newc\msdl{m_{\sdl}}
\newc\sdr{{\widetilde d}_R}      \newc\msdr{m_{\sdr}}

\newcommand\stopone{{\widetilde t}_1}   \newcommand\mstopone{m_{\stopone}}
\newcommand\stoptwo{{\widetilde t}_2}   \newcommand\mstoptwo{m_{\stoptwo}}

\newc\sfermion{\tilde f}  \newc\msfermion{m_{\sfermion}}

\newc\cmeter{{\rm cm}} \newc\meter{{\rm m}} \newc\kmeter{{\rm km}}

\newc\second{{\rm s}} 
\newc\sr{{\rm sr}}

\newc{\gstar}{g_\ast}           \newc{\gsstar}{g_{s\ast}}
\newc{\geff}{g_{\rm eff}}
\newcommand\mz{m_{Z}}

\newc{\sthw}{\sin\theta_W}              \newc{\cthw}{\cos\theta_W}
\newc{\bino}{\widetilde B}              \newc{\wino}{\widetilde W_30}
\newc{\higgsinob}{{\widetilde H}^0_b}   \newc{\higgsinot}{{\widetilde H}^0_t}
\newc{\abund}{\Omega h^2}
\newc{\abundchi}{\Omega_\chi h^2}
\newc{\abundcdm}{\Omega_{\text{CDM}} h^2}
\newc{\omegam}{\Omega_{M}}       \newc{\abundm}{\Omega_{M} h^2}
\newc{\omegab}{\Omega_{b}}       \newc{\abundb}{\Omega_{b} h^2}
\newc{\omegacdm}{\Omega_{CDM}}
\newc{\omegatot}{\Omega_{TOT}}
\newc{\rhocrit}{\rho_{crit}}
\newc{\rhochi}{\rho_{\chi}}
\newc\pb{\,\mbox{pb}} \newc\fb{\,\mbox{fb}}

\newc\pc{\,\mbox{pc}} \newc\kpc{\,\mbox{kpc}}
\newc\mpc{\,\mbox{Mpc}} \newc\gpc{\,\mbox{Gpc}}

\newc\BR{BR}
\newc\bsgamma{b\rightarrow s \gamma }
\newc\bxsgamma{\overline{B}\rightarrow X_{s}\gamma}
\newc\brbsgamma{\BR(\overline{B}\rightarrow X_s\gamma)}

\newcommand\brbsmumu{\BR(\overline{B}_s\to\mu^+\mu^-)}

\newcommand\delmbs{\Delta M_{B_s}}

\newc{\beq}{\begin{equation}}
\newc{\eeq}{\end{equation}}

\newc\stoponetwo{{\widetilde t}_{1,2}}
\newc\sbotonetwo{{\widetilde b}_{1,2}}
\newc\stauonetwo{{\widetilde \tau}_{1,2}}

\newc{\sigsip}{\sigma^{SI}_{p}} \newc{\sigsin}{\sigma^{SI}_{n}}
\newc{\sigsiN}{\sigma^{SI}_{N}}
\newc{\sigsdp}{\sigma^{SD}_{p}} \newc{\sigsdn}{\sigma^{SD}_{n}}
\newc{\sigsiA}{\sigma^{SI}_{A}}

\newc{\pbar}{\bar{p}}

\newc{\egamma}{E_{\gamma}}
\newc{\flux}[1]{\Phi_{#1}}
\newc{\dfluxde}[1]{\frac{d\Phi_{#1}}{d E_{#1}}}

\newc{\fluxg}{\Phi_{\gamma}}
\newc{\dfluxgde}{\frac{d\Phi_{\gamma}}{d\egamma}}
\newc{\dfluxgdetext}{ d\Phi_{\gamma} / d\egamma}

\newc{\eplus}{e^+}
\newc{\epos}{E_{\eplus}}
\newc{\eps}{\varepsilon}

\newc{\npos}{n_{\eplus}} \newc{\Npos}{N_{\eplus}}
\newc{\dnposde}{\frac{d n_{\eplus}}{d\epos}}
\newc{\dnposdeps}{\frac{d n_{\eplus}}{d\eps\phantom{_{\eplus}}}}
\newc{\dnposdepstext}{ d n_{\eplus} / d\eps}
\newc{\fluxpos}{\Phi_{\eplus}}  \newc{\fluxelec}{\Phi_{e^{-}}}
\newc{\dfluxposde}{\frac{d\Phi_{\eplus}}{d\epos}}
\newc{\dfluxposdetext}{ d\Phi_{\eplus} / d\epos}

\newc{\chisq}{\chi^2}  \newc{\chisqred}{\chi^2_{\text{red}}}

\newc{\nfwc}{{\text{NFW+ac}}} \newc{\moorec}{{\text{Moore+ac}}}

\newc\xilim{\xi_{\rm lim}} 
\newc\tlim{t_{\rm lim}} 
\newc\zetalim{\zeta_{\rm lim}} 

\newc\zetah{\zeta_h}
\newc{\relprobone}[1]{p({#1} \vert d)}
\newc{\relprobtwo}[2]{p({#1},{#2} \vert d)}

\long\def\begincomment#1\endcomment{%
        \begingroup\sf\baselineskip12pt#1\endgroup}

\newc\ANAP[3]
                 {{\em Astron. Astrophys.} {\bf #1} (#2) #3} 
\newc\AP[3]
                {{\em Ann. Phys.} {\bf #1} (#2) #3}
\newc\APJ[3]
                {{\em Astrophys. J.} {\bf #1} (#2) #3}
\newc\APP[3]
                {{\em Astropart. Phys.} {\bf #1} (#2) #3}
\newc\APS[3]
                {{\em Astrophys. J. Suppl.} {\bf #1} (#2) #3}

\newc\ARNPS[3]
                {{\em Ann. Rev. Nucl. Part. Sci.} {\bf C#1} (#2) #3}
\newc\CPC[3]
                {{\em Comp. Phys. Comm.} {\bf C#1} (#2) #3}
\newc\EPJ[3]
                {{\em Eur. Phys. J.} {\bf C#1} (#2) #3}
\newc\JCAP[3]
                {{\em JCAP} {\bf #1} (#2) #3}
\newc\JHEP[3]
                {{\em JHEP} {\bf #1} (#2) #3}
\newc\IJMP[3]
                {{\em Int. J. Mod. Phys.} {\bf A#1} (#2) #3}
\newc\MNRAS[3]
                 {{\em Mon. Not. Roy. Astron. Soc.} {\bf #1} (#2) #3}
\newc\MPL[3]
                {{\em Mod. Phys. Lett.} {\bf A#1} (#2) #3}
\newc\NCA[3]
                {{\em Nuovo Cimento} {\bf #1} (#2) #3}
\newc\NIM[3]
                {{\em Nucl. Instrum. Methods} {\bf #1} (#2) #3}
\newc\NIMA[3]
                {{\em Nucl. Instrum. Methods} {\bf A#1} (#2) #3}
\newc\NAT[3]
                {{\em Nature} {\bf #1} (#2) #3}
\newc\NPB[3]
                {{\em Nucl. Phys.} {\bf B#1} (#2) #3}
\newc\PL[3]
                {{\em Phys. Lett.} {\bf B#1} (#2) #3}
\newc\PLB[3]
                {{\em Phys. Lett.} {\bf B#1} (#2) #3}
\newc\PR[3]
                {{\em Phys. Rep.} {\bf #1} (#2) #3}
\newc\PRL[3]
                {{\em Phys. Rev. Lett.} {\bf #1} (#2) #3}
\newc\PRD[3]
                {{\em Phys. Rev.} {\bf D#1} (#2) #3}
\newc\PTP[3]
                {{\em Prog. Theor. Phys.} {\bf #1} (#2) #3}
\newc\PPNP[3]
                {{\em Prog. Part. Nucl. Phys.} {\bf #1} (#2) #3}
\newc\RMP[3]
                {{\em Rev. Mod. Phys.} {\bf #1} (#2) #3 }
\newc\RPP[3]
                {{\em Rept. Prog. Phys.} {\bf #1} (#2) #3 }
\newc\SC[3]
                {{\em Science} {\bf #1} (#2) #3 }
\newc\ZPC[3]
                {{\em Z. Phys.} {\bf C#1} (#2) #3}
\newc\Err[3]
           {{\em Erratum-ibid.} {\bf #1} (#2) #3 }

\newcommand{\squishlist}{
   \begin{list}{$\bullet$}
    { \setlength{\itemsep}{0pt}      \setlength{\parsep}{3pt}
      \setlength{\topsep}{3pt}       \setlength{\partopsep}{0pt}
      \setlength{\leftmargin}{1.em} \setlength{\labelwidth}{1em}
      \setlength{\labelsep}{0.5em} } }
\newcommand{\squishend}{
    \end{list}  }
        \newcommand\mW{m_{W}}

\newcommand{\sineff}{\sin^2 \theta_{\text{eff}}}

\begin{document}

\begin{titlepage}
\pagestyle{empty}
\baselineskip=21pt
\vskip 1.5cm
\begin{center}

{\Large\bf On prospects for dark matter indirect detection\\in the
  Constrained MSSM}

\end{center}   
\begin{center}   
\vskip 0.75 cm
{\bf Leszek Roszkowski}${}^a$\,\footnote{\, L.Roszkowski@sheffield.ac.uk}, 
{\bf Roberto Ruiz de Austri}${}^b$\,\footnote{\, rruiz@delta.ft.uam.es},
{\bf Joe Silk}${}^c$\,\footnote{\, silk@astro.ox.ac.uk} 
and {\bf Roberto Trotta}${}^c$\,\footnote{\, rxt@astro.ox.ac.uk}
\vskip 0.1in
\vskip 0.4cm ${}^a$ {\it Department of Physics and
Astronomy, University of Sheffield, Sheffield, S3 7RH, UK}\\ 
${}^b$ {\it Departamento de F\'{\i}sica Te\'{o}rica C-XI
        and Instituto de F\'{\i}sica Te\'{o}rica C-XVI,
        Universidad Aut\'{o}noma de Madrid, Cantoblanco,
        28049 Madrid, Spain}\\ 
${}^c$ {\it Oxford University, Department of Astrophysics,  Denys Wilkinson Building,
Keble Road,\\Oxford, OX1 3RH, UK}

\vskip 1cm
\abstract{ We apply a rigorous statistical analysis to the Constrained
MSSM to derive the most probable ranges of the diffuse gamma radiation
flux from the direction of the Galactic center and of the positron
flux from the Galactic halo due to neutralino dark matter
annihilation, for several different choices of the halo model and propagation
model parameters. We find that, for a specified halo profile, and
assuming flat priors, the 68\% probability range of the integrated
$\gamma$--ray flux spans about one order of magnitude, while the 95\%
probability range can be much larger and extend over four orders of
magnitude (even exceeding five for a tiny region at small
neutralino mass). The detectability of the signal by GLAST depending
primarily on the cuspiness of the halo profile. The positron flux, on
the other hand, appears to be too small to be detectable by PAMELA,
unless the boost factor is at least of order ten and/or the halo
profile is extremely cuspy.  We also briefly discuss the sensitivity
of our results to the choice of priors.}
\end{center}
\baselineskip=18pt \noindent

\vfill
\leftline{PACS: 12.60.Jv, 14.80.Ly, 95.35.+d}
\end{titlepage}

\section{Introduction}\label{sec:intro}
There is currently much evidence for the
existence of large amounts of dark matter (DM) in the Universe.
While its nature remains unknown, DM is likely to be made up of an
exotic species of weakly interacting massive particles (WIMPs). A
particularly popular WIMP candidate is the lightest neutralino
$\chi$ of effective low-energy supersymmetry (SUSY), which is
stable due to R-parity~\cite{susy-dm-reviews,bhs04}. In addition
to collider searches for SUSY and direct detection (DD) searches
for a cosmic WIMP, a promising strategy is that of indirect detection
(ID), \ie, a search for traces of WIMP pair-annihilation in
the Milky Way. Since the annihilation rate is proportional to the
square of the WIMP number density, of particular interest are the
Galactic center (GC) and nearby clumps in the halo where the
density of DM is believed to be enhanced. The aim of this paper is
to provide, for the first time, a statistical measure for the
prediction of $\gamma$--ray and positron signatures in low-energy
SUSY, thus allowing one to assess high-probability regions for
DM-annihilation signatures that could be observed by the GLAST
(in orbit since June  2008) and PAMELA (launched 2006) satellites.
Existing data from EGRET suggest a spectrally distinct excess of
$\gamma$--rays up to $\sim 10\gev$ and the HEAT data indicate a
possible excess in positron flux between $5$ to $\sim 30\gev$.
GLAST and PAMELA will provide an order of magnitude more
sensitivity.

In assessing detection prospects of WIMPs there are two main
sources of uncertainties.  One comes from the underlying particle
physics model where WIMP mass and annihilation cross section can
vary over a few orders of magnitude. The other is astrophysical in
nature and stems from substantial uncertainties in the DM
distribution, both locally (local DM density and the existence of
clumps) and towards the GC.  Since the general Minimal
Supersymmetric Standard Model (MSSM) suffers from a lack of
predictability due to a large number of free parameters, it is
interesting and worthwhile to assess WIMP detection prospects in
more constrained and more well-motivated low-energy SUSY models,
among which particularly popular is the Constrained MSSM
(CMSSM)~\cite{kkrw94}, which includes the minimal supergravity
model~\cite{sugra-reviews}. By applying a statistical approach, we
derive in the CMSSM most probable ranges of fluxes, thus bringing
under control all the uncertainties of the particle physics side
of WIMP detection. This is a major improvement over existing
methods which are usually limited to the consideration of a few
representative choices of points or slices in the parameter space.
Detection prospects then become a function of specific
astrophysical uncertainties only.

In this Letter we employ a Bayesian Markov Chain Monte Carlo
(MCMC) technique to efficiently explore the multi-dimensional
parameter space of the CMSSM, and to include all relevant sources of
uncertainty on the particle physics side~\cite{rrt,rrt3} (for a similar
study, see~\cite{allanach-bayes}).  Our Bayesian approach allows us to
produce probability maps for all relevant observable quantities, thus
establishing a complete set of predictions of the CMSSM.

\section{Bayesian analysis of the CMSSM}\label{sec:bayesiancmssm}
The CMSSM is described in
terms of four free parameters: a ratio of Higgs vacuum
expectation values $\tanb$, and common soft SUSY-breaking mass
parameters of gauginos, $\mhalf$, scalars, $\mzero$, and tri-linear
couplings, $\azero$. The parameters $\mhalf$, $\mzero$ and
$\azero$ are specified at the GUT scale, $\mgut\simeq 2\times
10^{16}\gev$, which serves as a starting point for evolving the MSSM
renormalization group equations for
couplings and masses down to a low energy scale
$\msusy\equiv \sqrt{\mstopone\mstoptwo}$ (where
$m_{\stopone,\stoptwo}$ denote the masses of the scalar partners of
the top quark), chosen so as to minimize higher order loop
corrections. At $\msusy$ the (1-loop corrected) conditions of
electroweak symmetry breaking (EWSB) are imposed. The sign of the
Higgs/higgsino mass parameter $\mu$, however, remains
undetermined. Here we set $\mu>0$.

In deriving predictions for the observable quantities, one also needs
to take into account the uncertainty coming from our imperfect
knowledge of the values of some relevant Standard Model (SM)
parameters, namely the pole top quark mass, $\mtpole$, the bottom
quark mass at $m_b$, $\mbmbmsbar$, and the electromagnetic and the
strong coupling constants at the $Z$ pole mass $M_Z$, $\alphaemmz$ and
$\alphas$, respectively (the last three quantities are all computed in
the $\msbar$ scheme). These four ``nuisance parameters'' are the most
relevant ones for accurately predicting the SUSY spectrum and its
observable signature. In our analysis we thus consider an
8-dimensional parameter space spanned by the above four SM and the
four CMSSM parameters. 

In general, the results of a Bayesian analysis are expressed
in terms of a posterior probability distribution (or more briefly,
``a posterior''). By virtue of Bayes' theorem, the posterior is the
product of the prior and the likelihood. The prior expresses the state
of knowledge about the parameters before seeing the data, while the
likelihood encodes the information coming from the observations (for
further details on the Bayesian framework, see
e.g.~\cite{Trotta:2008qt}). If the constraining power of the data is
strong enough, then the posterior is effectively dominated by the
likelihood and the prior distribution becomes irrelevant. However, if
the observations are not sufficiently constraining, the posterior will
retain a prior dependence. Therefore it is important to check to which
extent the results based on the posterior pdf show a prior
dependency. There are reasons to believe that for the CMSSM present
data are not sufficiently powerful to completely override the prior,
see~\cite{tfhrr1} for a detailed study of this issue.

In our analysis we assume flat priors on both
SM and CMSSM parameters over wide ranges of their values, encompassing
the focus point region~\cite{focuspoint-orig}. However, below we will
comment on how our result change when one employs a flat prior in
$\log_{10}\mhalf$ and $\log_{10}\mzero$ instead (which we call in the
following ``the log prior'' for brevity). The reason for this
alternative choice of prior is that that they are distinctively
different.  In particular, the log prior gives equal {\em a priori}
weights to all decades for the parameters. So the log prior expands
the low-mass region and allows a much more refined scan in the
parameter space region where finely tuned points can give a good fit
to the data (see~\cite{tfhrr1} for details).  Other choices of priors
are possible and indeed physically motivated, and will be considered
in future work. A recent discussion of some alternative prior choices
in the CMSSM (motivated by considerations of naturalness and fine
tuning) can be found in Ref.~\cite{allanach06}.

At every point in parameter space, we compute a number of observable
quantities, and compare their values with the observational data
listed in Ref.~\cite{rrt3},\footnote{We employ the WMAP 3-year relic
abundance value assuming that neutralinos are the only dark matter
component. Using the WMAP 5-year value instead would not change
visibly our results.} where also a detailed description of our
procedure is given.  We include all relevant collider limits,
including direct limits on Higgs and superpartner masses, rare
processes $\brbsgamma$, $\brbsmumu$ and recently measured $B_s$
mixing, $\delmbs$, electroweak precision data ($\mW$ and $\sineff$)
and the relic abundance of the lightest neutralino $\abundchi$ assumed
to be the cold DM in the Universe. We then use our MCMC
algorithm~\cite{sb} to produce, for a given model of DM distribution
in the Galactic halo, probability distribution maps in parameter space
and various observables, including ID ones which are computed with the
help of DarkSusy~\cite{darksusy}. As we have emphasized in
Ref.~\cite{rrt3}, current constraints, especially from $\bsgamma$
favor the focus point region of large $\mzero\gsim1\tev$ and not so
large $\mhalf\lsim1.5\tev$ (with $\mhalf\lsim2.5\mzero$).

\section{Gamma-ray flux from the Galactic center}\label{sec:gammaflux}
The
differential diffuse $\gamma$--ray flux arriving from a direction at
an angle $\psi$ from the GC is given by~\cite{bhs04}
\beq
\dfluxgde (\egamma, \psi) = \sum_{i} \frac{\sigma_i
v}{8\pi\mchi^2}\, \frac{d N^i_\gamma}{d\egamma}
\int_{\text{l.o.s.}} dl\, \rhochi^2(r(l,\psi)),
\label{eq:diffgammaflux}
\eeq
where $\sigma_i v$ is a product of the WIMP pair-annihilation
cross section into a final state $i$ times the pair's relative
velocity and $d N^i_\gamma /d\egamma$ is the differential
$\gamma$--ray spectrum (including a branching ratio into photons)
following from the state $i$. Here we consider contributions from
the continuum (as opposed to photon lines coming from one loop
direct neutralino annihilation into $\gamma\gamma$ and $\gamma
Z$), resulting from cascade decays of all kinematically allowed final state
SM fermions and combinations of gauge and Higgs bosons. The integral
is taken along the line of sight (l.o.s.) from the detector. It is
convenient to separate factors depending on particle physics and
on halo properties by introducing the dimensionless quantity
$J(\psi)\equiv \left(1/8.5\kpc\right)\left(
0.3\gev/\cmeter^3\right)^2 \int_{\text{l.o.s.}} dl\,
\rhochi^2(r(l,\psi))$~\cite{bub97}. The flux is further averaged
over the solid angle $\Delta\Omega$ representing the acceptance
angle of the detector, and one defines the quantity ${\bar
J}(\Delta\Omega)=\left(1/\Delta\Omega\right) \int_{\Delta\Omega}
J(\psi) d\Omega$. 

Clearly, one of the crucial ingredients is the
radial dependence of the WIMP density $\rhochi(r)$. Some popular
profiles can be parameterized by~\cite{bhs04} 
\begin{equation}
\rhochi(r)= \rho_0 \frac{(r/r_0)^{-\gamma}}
            {\left[1+\left(r/a\right)^\alpha\right]^{\frac{\beta-\gamma}{\alpha}}}
             \left[1+(r_0/a)^\alpha\right]^{\frac{\beta-\gamma}{\alpha}},
\label{eq:halomodels}
\end{equation}
where the halo WIMP density has been normalized to its local
value, assumed to be $\rho_0=0.3\gev/\cmeter^3$.
Table~\ref{tab:halomodels} gives the values of the parameters: $a$,
$\alpha$, $\beta$, $\gamma$ and $r_0$ for some common
choices. Here we consider the line-of-sight (l.o.s.)
integration factor $\bar{J}$ in the direction of the GC, \ie, for
$\psi=0$. In the case of the cuspy profiles, in order to avoid a
divergent behavior, we set a cutoff radius of $r_c=10^{-5}\kpc$.
\begin{table}
\centering
\begin{tabular}{| c | c | c c c | c | c |}
\hline
 Halo model & $a$ ($\kpc$)  &$\alpha$ &$\beta$
 &$\gamma$ & ${\bar J}(10^{-3}\sr)$ & ${\bar J}(10^{-5}\sr)$\\ \hline\hline
 isothermal cored    & 3.5   & 2 & 2 & 0 & 30.35 & 30.40 \\ \hline
 NFW & 20.0  & 1 & 3 & 1 & $1.21 \times 10^3$ & $1.26 \times 10^4$ \\ \hline
 $\nfwc$ & 20.0 & 0.8 & 2.7 &  1.45 &  $1.25 \times 10^5$ &  $1.02 \times 10^7$ \\ \hline
 Moore   & 28.0 & 1.5 & 3 & 1.5 &  $1.05 \times 10^5 $ &  $9.68 \times 10^6 $ \\ \hline
 $\moorec$& 28.0& 0.8 & 2.7 & 1.65 & $1.59 \times 10^6$ & $3.12 \times 10^8$ \\ \hline
\end{tabular}
\caption{Parameters for some popular halo profiles: a spherically
  symmetric modified isothermal model~\cite{isothermalmodel},
  the Navarro, Frenk and White (NFW) model~\cite{nfwhalo95} and the
  Moore, \etal, (Moore) model~\cite{morehalo99}. Everywhere $r_0 = 8.0\kpc$
  except for the isothermal case, where $r_0 = 8.5 \kpc$. In the NFW
  and Moore, \etal, models
  the effect of adiabatic compression due to baryons (marked with an
  additional $\rm +ac$), is included. See also Ref.~\cite{mmnp05}.
\label{tab:halomodels}}
\end{table}
The total $\gamma$--ray flux from the cone $\Delta\Omega$ centered on
$\psi$ and integrated over photon energy from an energy threshold
$E_{\text{th}}$, is then given by
\beq
\fluxg(\Delta\Omega) =  \int^{\mchi}_{E_{\text{th}}}
d\egamma\, \dfluxgdetext(\egamma, \Delta\Omega).
\label{eq:totalgacflux}
\eeq

\begin{figure}[htb!]
\centering
\includegraphics[width=0.6\textwidth]{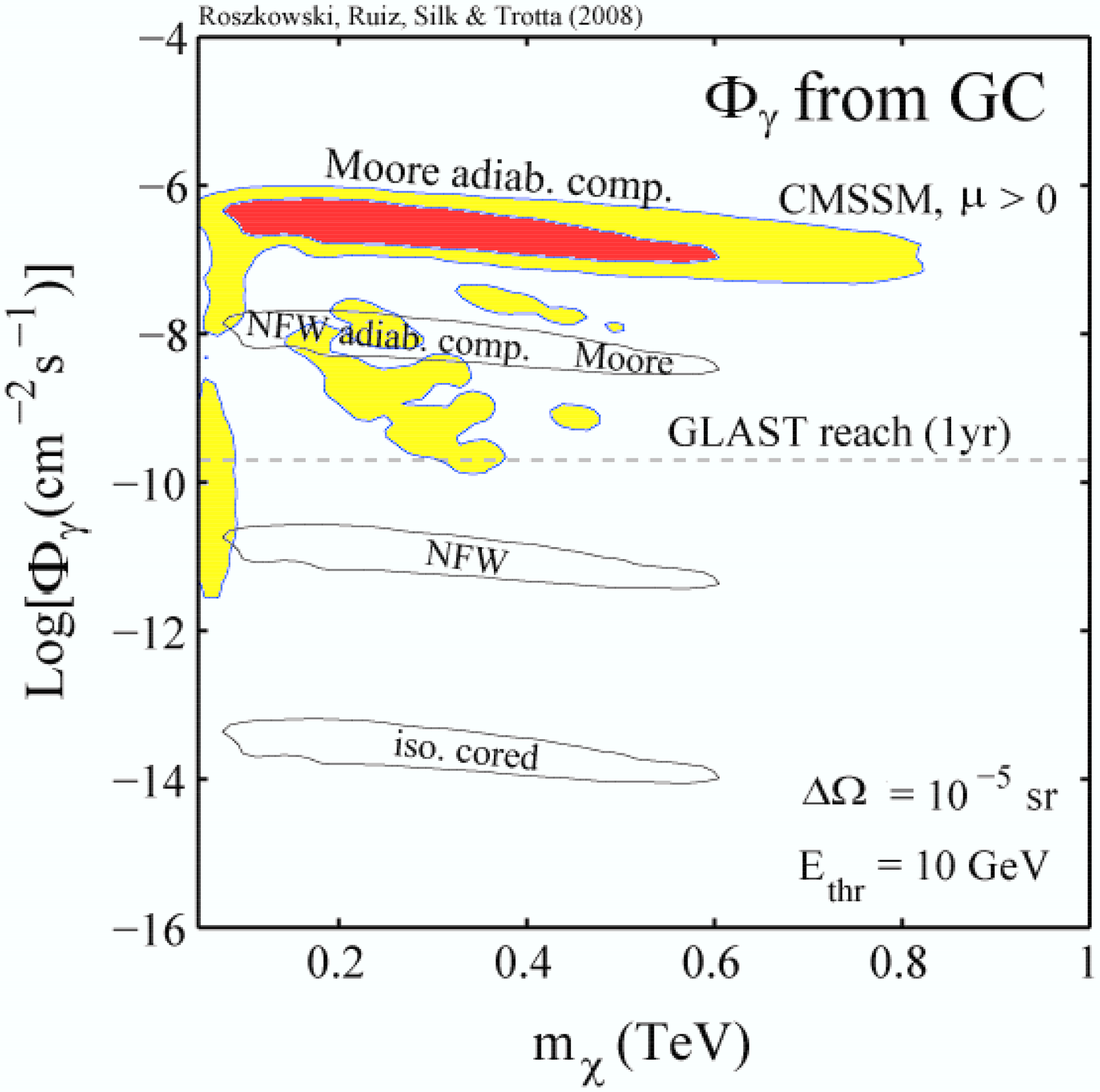}\\
\caption{The joint probability distribution for the $\gamma$--ray flux
  $\fluxg$ from the Galactic center vs the neutralino mass $\mchi$ for
  some popular halo profile models, assuming flat priors, as explained
  in the text. The dark/red (light/yellow) region
  shows the predicted 68\% (95\%) probability ranges for the Moore
  profile with adiabatic compression (for other profiles we indicate
  only the limits of the 68\% region).  We also plot the expected
  5-$\sigma$ detection threshold (neglecting background) for
  energies above $E_\gamma = 10\gev$ for GLAST after 1 year of
  operation~\cite{glast-reach-own}. Notice that for GLAST to be able
  to detect the annihilation flux over the background one might require
  much larger fluxes than the sensitivity level plotted
  here~\cite{Bertone:2006kr}.  \label{fig:grfluxvsmchi}}
\end{figure}

With the launch of GLAST, which has angular
resolution $\Delta\Omega\simeq10^{-5}\sr$ and sensitivity to
fluxes larger than about $2\times10^{-10}\cmeter^{-2}\second^{-1}$
for photon energies $E_\gamma \gsim
10\gev$~\cite{glast-reach-own},\footnote{Resolution and 
sensitivity in the range $30\mev \lsim E_\gamma \lsim 10 \gev$ are
energy-dependent and would require a more careful
analysis.} it is timely to investigate the
global predictions of the CMSSM for a range of halo models.

Fig.~\ref{fig:grfluxvsmchi} shows the joint probability
distribution for the total flux $\fluxg$ from the GC above a
threshold energy of $10\gev$ versus $\mchi$, integrated over
$\Delta\Omega = 10^{-5}\sr$. The spread of values reflects the
marginalization over all the four CMSSM and four SM parameters,
thus fully accounting for all substantial sources of uncertainty on the
particle physics side. Firstly, as can be seen from the figure, the
2-dim joint 
68\% probability range of $\mchi$ lies between about $80\gev$ and about
$600\gev$. Secondly, for a given halo profile, and
assuming flat priors in CMSSM parameters, we find that the
68\% probability range of $\fluxg$ is confined to lie within about
one order of magnitude. On the other hand, the spread of the 95\%
probability range is much larger  and at lower $\mchi$ can extend over
four or even five orders of magnitude. 

In order to examine the low mass region in more detail, we have redone
our analysis for the log prior choice introduced above. As $\mhalf$
and $\mzero$ are the primary CMSSM parameters determining mass spectra
of the neutralino, the other superpartners and the Higgs bosons, the
log prior allows one to examine the low mass region in more detail, in
particular by ``expanding'' the volume of the region $100\gev
\lsim\mhalf,\mzero \lsim 1\tev$. As we discuss below, the flat prior
appears to produce an optimistic scenario as far as indirect detection
signatures are concerned, while the log prior can give lower values of
the fluxes and hence it leads to more pessimistic prospects for
indirect detection. Ways of mediating between the two scenarios and to
assess their relative plausibility will be explored in future work.
 
Since the log prior gives more ``weight'' to lower values of both
$\mhalf$ and $\mzero$, not surprisingly, we have found that it leads
to a large widening of mostly the lower boundary of the 68\%
probability range at low $\mchi$, while not affecting the flux ranges
at larger values of the neutralino mass. For example, the 68\%
probability range widens to nearly three decades and, in the case of
the Moore profile with adiabatic compression, can be as low as
$1.2\times10^{-10}\cmeter^{-2}\second^{-1}$ at $\mchi\sim100\gev$, but
then it quickly raises and for $\mchi\gsim200\gev$ is not very
different from the case of the flat prior. A more detailed discussion
of the implications for CMSSM parameters of employing a log prior is given in Ref.~\cite{tfhrr1}.

For a given prior, choosing a different halo profile merely amounts to
shifting the total flux by the ratio of the values of ${\bar J}$ given
in Table~\ref{tab:halomodels}.  As expected, more cuspy profiles lead
to higher predicted fluxes.  We find that, in the CMSSM in the case of
the Moore profile (with and without adiabatic compression) and the NFW
profile with adiabatic compression, the continuum flux signal will be
within the reach of GLAST, while for profiles with
$\bar{J}(10^{-5}\sr) \lsim 10^5$ it will not be detectable by
GLAST. (We have checked that the case of $\mu<0$ and flat priors gives
qualitatively similar results.)

\begin{figure}[t!]
\centering
\includegraphics[width=0.6\textwidth]{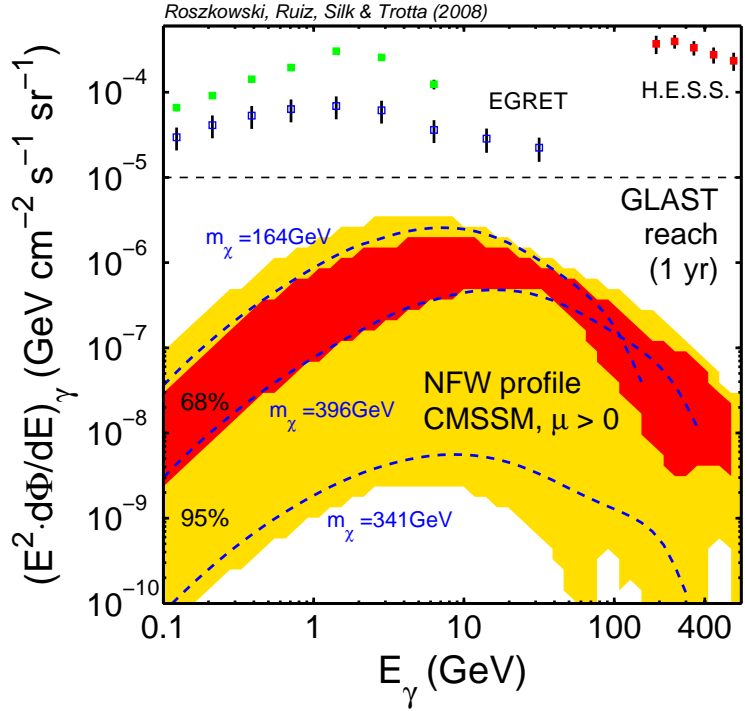}
\caption{Predicted $\gamma$--ray differential energy flux averaged
over a solid angle $\Delta \Omega = 10^{-3}\sr$ and fully
accounting for current uncertainty in the CMSSM parameters, assuming flat priors. The
$68\%$ and $95\%$ regions are for the NFW profile, all other cases
can be obtained by rescaling them by the factors $\bar{J}$ given
in Table~\ref{tab:halomodels}. Predictions appropriate for GLAST
resolution ($\Delta \Omega = 10^{-5}\sr$) are obtained by dividing
by $1.21 \times 10^3$ and multiplying by the desired value of
$\bar{J}(10^{-5}\sr)$. We plot the expected GLAST $8\sigma$
detection threshold (horizontal black/dashed
line~\cite{glast-reach-own}). The three blue, dashed curves show
sample spectra (for the values of $\mchi$ specified in the figure) from our 
statistical scan. For comparison we plot EGRET diffuse data
towards the GC (green
squares~\cite{egret-excessdata}), EGRET's point-source subtracted
flux (blue empty squares~\cite{strong}) and H.E.S.S. (2004)
data~\cite{hess04} (red squares) with $2\sigma$ error bars.}
\label{fig:diffgrfluxvsmchi}
\end{figure}

The differential $\gamma$--ray flux from DM annihilations is expected
to exhibit a sharp drop-off in the energy spectrum as $\egamma$
approaches $\mchi$. In Fig.~\ref{fig:diffgrfluxvsmchi} we plot $68\%$
and $95\%$ probability regions for the $\gamma$--ray differential flux
for the NFW profile, averaged over a solid angle $\Delta \Omega =
10^{-3}\sr$ (to allow a comparison with EGRET data), for the flat prior choice.  Clearly, the
current uncertainty on CMSSM parameters and hence on $\mchi$
introduces a considerable spread in the predicted spectral shape of
the signal. Additional uncertainty comes from the dependence on the
priors. For example, for the log prior given above the 68\%
probability range of the differential photon range extends between
$2.4\times10^{-11}\gev\cmeter^{-2}\second^{-1}\sr^{-1}$ and
$6.7\times10^{-7}\gev\cmeter^{-2}\second^{-1}\sr^{-1}$.  Thus, even if
a positive signal were detected by GLAST, it would be difficult to
infer from it the mass of the WIMP, especially at its lower values
below some $200\gev$, with any reasonable accuracy.

\section{Positron flux from the Galactic halo}\label{sec:positronflux}
\begin{figure}[tb!]
\centering
\includegraphics[width=0.6\textwidth]{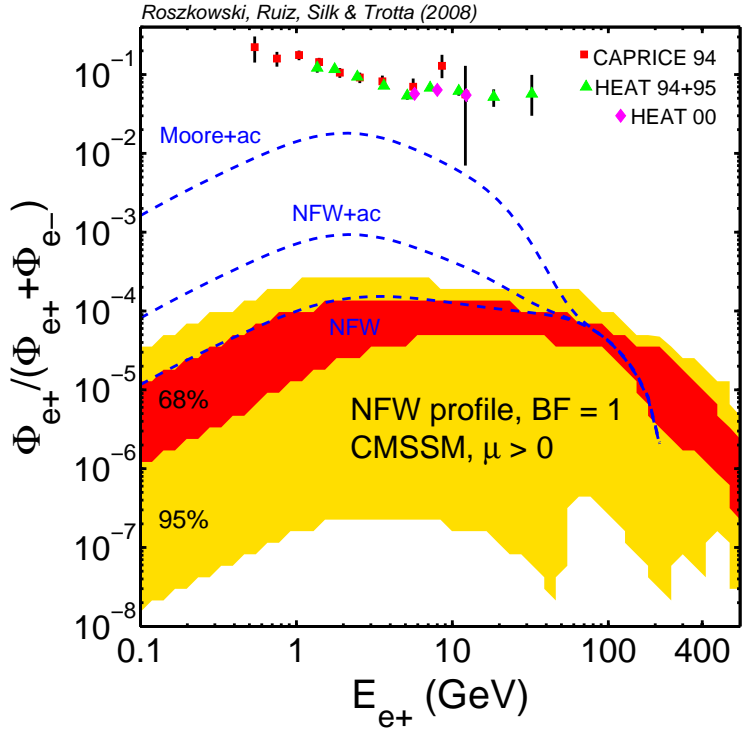} 
\caption{Predicted positron flux fraction in the CMSSM. The $68\%$
(dark/red) and $95\%$ (light/yellow) regions are for an NFW
profile with a boost factor BF=1 and a specific choice of
propagation model. We also show for comparison some of the current
data. To illustrate the dependency of the spectral shape at low
energies on the halo model, we plot the spectrum for the same
choice of CMSSM parameters (with $\mchi=229\gev$) for three
different halo models as indicated. In absence of a large boost
factor, the signal appears too small to be detected by PAMELA.}
\label{fig:diffposfluxvsepos}
\end{figure}

Positrons can be
produced either in direct DM annihilation, or from decays and
hadronization of other products (gauge and Higgs bosons, etc),
with the continuum spectrum from the latter usually dominating.
Once produced, they propagate through the Galactic medium and
their spectrum is distorted due to synchrotron radiation and
inverse Compton scattering at large energies, bremsstrahlung and
ionization at lower energies. The effects of positron propagation
are computed  following a standard procedure described 
in~\cite{esu04,be98}, by solving numerically the
diffusion-loss equation for the number density of positrons per
unit energy $\dnposdepstext$. The diffusion coefficient is
parameterized as $K(\eps) = K_0(3^\alpha + \eps^{\alpha})$, with
$K_0 = 5.8 \times 10^{27}~\cmeter^{2}\second^{-1}$, $\alpha = 0.6$
and $\eps= \epos/1\gev$, mimicking re-acceleration effects. The
energy loss rate is given by $b(\eps) = \tau_E\eps^2$, with
$\tau_E = 10^{-16}\second^{-1}$, and we describe the diffusion
zone (\ie, the Galaxy) as an infinite slab of height $L=4\kpc$,
with free escape boundary conditions. Changes in the above
positron propagation model, especially $K(\eps)$ (see e.g.
\cite{ms98,be98}), can potentially lead to variations by a factor
of 5 to 10 in the spectral shape at low positron energy, $\epos
\lsim 20\gev$~\cite{hs04}. In this energy region the flux
dependence on the halo profile is also substantial and, for the
models in Table~\ref{tab:halomodels}, the flux can change by up to
a few orders of magnitude (compare blue/dashed lines in
Fig.~\ref{fig:diffposfluxvsepos}), since positrons from further
away loose energy due to propagation.  Most
high-energy positrons, on the other hand, originate from the local
neighbourhood the size of a few~$\kpc$s~\cite{be98,lpst06}, and their
flux is less dependent of the halo and propagation dynamics. The
flux can, however, be considerably enhanced by the presence of local
DM clumps that survive merging processes and tidal
stripping~\cite{Diemand:2006ik}, an effect that is usually
parameterized by a boost factor (BF), which
can be of order 10. Recent studies have begun investigating the
clumpiness dependence of the spectrum in more
detail~\cite{cs06,lpst06}. Finally, in order to reduce the impact
of solar winds and magnetosphere effects on the model's predictions,
it is useful to consider the positron fraction, defined as
$\fluxpos/(\fluxpos + \fluxelec)$, where $\fluxpos$ is the
positron differential flux from WIMP
annihilation, while $\fluxelec$ is the
background electron flux. For background $e^-$ and $e^+$ fluxes we
follow the parametrization adopted in Ref.~\cite{be98} from
Ref.~\cite{ms98}.

In Fig.~\ref{fig:diffposfluxvsepos} we show the predicted positron
flux fraction in the CMSSM (for flat priors) for the NFW profile and a boost factor
BF=1, alongside a compilation of observations, most notably from
HEAT. Again, the uncertainty in the spectral shape is one of the main
results of our analysis, which accounts for the current uncertainties
regarding the CMSSM parameters. The 95\% probability region peaks in
the range $1\gev \lsim \epos \lsim 10\gev$, roughly in the region of
the apparent HEAT positron excess, but the strength of the signal is
insufficient for it to be detectable by PAMELA in the absence of a
large boost factor~\cite{mmn06}. On the other hand, in that energy
range the signal would be enhanced by more than one (two) order(s) of
magnitude for a more cuspy profile such as the NFW (Moore) profile
with adiabatic compression, as indicated in
Fig.~\ref{fig:diffposfluxvsepos} for a sample spectrum corresponding
to $\mchi=229\gev$. This is because in the case of more cuspy profiles
more high-energy electrons coming from the GC are scattered to lower
energies.  For the case of the log prior on $\mhalf$ and $\mzero$
we again find a significant decrease of the lower boundary of the 68\%
range. For example, for the NFW profile with BF=1 the ratio in
Fig.~\ref{fig:diffposfluxvsepos} can be as low as $3.8\times 10^{-9}$.

We
conclude that, for not exceedingly cuspy halo models, PAMELA is
unlikely to be sensitive to positron fluxes in the CMSSM, since
for the NFW profile the signal is more than two order of magnitude
smaller than the background. This result would qualitatively hold
even when taking into account the considerable uncertainties
coming from the boost factor due to local clumps and changes in
the positrons propagation model, each of which can potentially
change the spectrum by up to a factor of 10.


\section{Summary}\label{sec:summary}

In the framework of the Constrained MSSM, we have performed a Bayesian
analysis of prospects for indirect dark matter
detection via a diffuse $\gamma$--ray signal or a positron flux from
the Galactic center. 
This has allowed us to provide a statistically rigorous assessment
of the uncertainty from the particle physics side of the problem.

We found that the prospects for GLAST to detect a diffuse
$\gamma$--ray signal from the Galactic center depend primarily on the
cuspiness of the DM profile at small radii. For the choice of flat
priors in the CMSSM parameters, the NFW model appears to be a
borderline case, while a more cuspy halo would guarantee a signal for
a 68\% range of the CMSSM parameter space, except near the bottom end
of the neutralino mass around $100\gev$, below the 68\% probability
range of $\mchi$. In the low mass region the sensitivity to the choice
of priors remains however substantial. Adopting a log prior on
$\mhalf$ and $\mzero$ leads to a significant decreasing of the lower
boundary of the 68\% probability range of the $\gamma$--ray flux
towards lower values at low $\mchi\lsim200\gev$, but at larger $\mchi$
gives similar results as with flat priors.

On the other hand, a positron flux is unlikely to
be detectable by PAMELA for both choices of priors, unless it is
strongly enhanced by a nearby clump with a boost factor of at least of
order ten. The latter conclusion is valid for a specific (although well motivated) choice of propagation model parameters. 
Assumptions regarding propagation parameters could however be easily relaxed in our framework. It would be straightforward
to extend our treatment to include propagation model parameters as
nuisance parameters and marginalize over them, as well. It is expected that such a procedure would increase the present, very substantial uncertainty as to the spectral shape, which we have shown is a consequence of the current lack of knowledge as to the preferred regions of the CMSSM parameters. Finally, it would also be interesting to repeat this analysis in a more general phenomenological SUSY model than the Constrained MSSM. While a richer phenomenology might help in explaining future signals should they be detected, it is also clear that a larger number of free parameters on the particle physics side will add to the difficulty of reliably predicting the shape and strength of both the $\gamma$--ray and the positron spectra.

{\bf Acknowledgements} \\ 
We thank G.~Bertone, I.~Moskalenko and A.~Strong
for useful comments. R.RdA is supported by the program
``Juan de la Cierva'' of the Ministerio de Educaci\'{o}n y Ciencia of
Spain. RT is supported by the Royal Astronomical Society and St Anne's
College, Oxford. We acknowledge partial support from ENTApP, part of
ILIAS, and UniverseNet.



\begin{thebibliography}{99}
\bibitem{sb}
The code is available from {\texttt{superbayes.org}}.
\end{thebibliography}
\end{document}